\documentclass[conference,letterpaper,english,final]{IEEEtran}

\usepackage[nodisplayskipstretch]{setspace}
\usepackage[style=ieee, doi=false, isbn=false, url=false]{biblatex}
\usepackage{bm,bbm}
\usepackage{amsthm,amssymb,amsmath,mathtools}
\usepackage{microtype}
\usepackage{algpseudocode,algorithm,algorithmicx}
\usepackage[svgnames]{xcolor}
\usepackage{graphicx,tabularx,float}
\usepackage{subcaption}
\usepackage{babel,csquotes}
\usepackage{hyperref}

\makeatletter

\theoremstyle{plain}
\newtheorem{theorem}{Theorem}
\newtheorem{definition}[theorem]{Definition}
\newtheorem{lemma}[theorem]{Lemma}

\newtheorem{corollary}[theorem]{Corollary}

\bmdefine{\bA}{A}
\bmdefine{\ba}{a}
\bmdefine{\bB}{B}
\bmdefine{\bb}{b}
\bmdefine{\bC}{C}
\bmdefine{\bc}{c}
\bmdefine{\bD}{D}
\bmdefine{\bd}{d}
\bmdefine{\bE}{E}
\bmdefine{\be}{e}
\bmdefine{\bF}{F}
\bmdefine{\bf}{f}
\bmdefine{\bG}{G}
\bmdefine{\bg}{g}
\bmdefine{\bH}{H}
\bmdefine{\bh}{h}
\bmdefine{\bI}{I}
\bmdefine{\bi}{i}
\bmdefine{\bJ}{J}
\bmdefine{\bj}{j}
\bmdefine{\bK}{K}
\bmdefine{\bk}{k}
\bmdefine{\bL}{L}
\bmdefine{\bl}{l}
\bmdefine{\bM}{M}
\bmdefine{\bmm}{m}
\bmdefine{\bN}{N}
\bmdefine{\bn}{n}
\bmdefine{\bO}{O}
\bmdefine{\bo}{o}
\bmdefine{\bP}{P}
\bmdefine{\bp}{p}
\bmdefine{\bQ}{Q}
\bmdefine{\bq}{q}
\bmdefine{\bR}{R}
\bmdefine{\br}{r}
\bmdefine{\bS}{S}
\bmdefine{\bs}{s}
\bmdefine{\bT}{T}
\bmdefine{\bt}{t}
\bmdefine{\bU}{U}
\bmdefine{\bu}{u}
\bmdefine{\bV}{V}
\bmdefine{\bv}{v}
\bmdefine{\bW}{W}
\bmdefine{\bw}{w}
\bmdefine{\bX}{X}
\bmdefine{\bx}{x}
\bmdefine{\bY}{Y}
\bmdefine{\by}{y}
\bmdefine{\bZ}{Z}
\bmdefine{\bz}{z}

\bmdefine{\balpha}{\alpha}
\bmdefine{\bbeta}{\beta}
\bmdefine{\bgamma}{\gamma}
\bmdefine{\bGamma}{\Gamma}
\bmdefine{\bdelta}{\delta}
\bmdefine{\bDelta}{\Delta}
\bmdefine{\bepsilon}{\epsilon}
\bmdefine{\bvarepsilon}{\varepsilon}
\bmdefine{\bzeta}{\zeta}
\bmdefine{\bmeta}{\eta}
\bmdefine{\btheta}{\theta}
\bmdefine{\bTheta}{\Theta}
\bmdefine{\biota}{\iota}
\bmdefine{\bkappa}{\kappa}
\bmdefine{\blambda}{\lambda}
\bmdefine{\bLambda}{\Lambda}
\bmdefine{\bmu}{\mu}
\bmdefine{\bnu}{\nu}
\bmdefine{\bpi}{\pi}
\bmdefine{\bPi}{\Pi}
\bmdefine{\brho}{\rho}
\bmdefine{\bsigma}{\sigma}
\bmdefine{\bSigma}{\Sigma}
\bmdefine{\btau}{\tau}
\bmdefine{\bupsilon}{\upsilon}
\bmdefine{\bUpsilon}{\Upsilon}
\bmdefine{\bphi}{\phi}
\bmdefine{\bPhi}{\Phi}
\bmdefine{\bchi}{\chi}
\bmdefine{\bpsi}{\psi}
\bmdefine{\bPsi}{\Psi}
\bmdefine{\bomega}{\omega}
\bmdefine{\bOmage}{\Omega}

\newcommand{\bfJ}{\mathbf{J}}

\newcommand{\tautil}{\tilde{\tau}}
\newcommand{\btautil}{\tilde{\btau}}

\newcommand{\bctil}{\tilde{\bc}}

\newcommand{\butil}{\tilde{\bu}}

\newcommand{\bbC}{\mathbb{C}}

\newcommand{\bbR}{\mathbb{R}}
\newcommand{\bbT}{\mathbb{T}}
\newcommand{\bbZ}{\mathbb{Z}}

\DeclareMathOperator{\diag}{diag}

\let\tilde\widetilde

\newcommand*{\herm}{{\mathsf{H}}}
\bmdefine{\bWW}{WW}

\newcommand{\pprime}{{\prime\prime}}

\newcommand\norm[1]{\left\lVert #1 \right\rVert}

\newcommand{\ellp}{{\ell^\prime}}

\makeatother

\allowdisplaybreaks[1]

\hypersetup{
    pdftitle={Physical Layer Location Privacy in SIMO Communication Using Fake Path Injection},
    pdfauthor={Tran Trong Duy, Maxime Ferreira Da Costa, Nguyen Linh Trung},
    pdfsubject={Information Theory, Physical Layer Security}
}

\title{Physical Layer Location Privacy in SIMO Communication Using Fake Path Injection}

\IEEEoverridecommandlockouts

\author{
  \IEEEauthorblockN{Tran Trong Duy\IEEEauthorrefmark{1}\IEEEauthorrefmark{2},
                    Maxime Ferreira Da Costa\IEEEauthorrefmark{2},
                    and Nguyen Linh Trung\IEEEauthorrefmark{1}
}
  \IEEEauthorblockA{\IEEEauthorrefmark{1}%
                   AVITECH, VNU University of Engineering \& Technology, Hanoi, Vietnam}
  \IEEEauthorblockA{\IEEEauthorrefmark{2}%
                    CentraleSupélec, Université Paris--Saclay, CNRS, Laboratory of Signals and Systems, Gif-sur-Yvette, France}
    \thanks{Tran Trong Duy's work is funded by the Master, PhD Scholarship Programme of Vingroup Innovation Foundation (VINIF): VINIF.2022.ThS.018.}
    \thanks{M. Ferreira Da Costa's work is supported by ANR-20-IDES-0002 and ANR-24-CE48-3094 grants.}
}

\begin{document}

\maketitle

\begin{abstract}
Fake path injection is an emerging paradigm for inducing privacy over wireless networks. In this paper, fake paths are injected by the transmitters into a single-input multiple-output (SIMO) communication channel to obscure their physical location from an eavesdropper. The case where the receiver (Bob) and the eavesdropper (Eve) use a linear uniform array to locate the transmitter's (Alice) position is considered. A novel statistical privacy metric is defined as the ratio between the smallest (resp. largest) eigenvalues of Eve's (resp. Bob's) Cramér-Rao lower bound (CRB) on the SIMO channel parameters to assess the privacy enhancements. Leveraging the spectral properties of generalized Vandermonde matrices, bounds on the privacy margin of the proposed scheme are derived. Specifically, it is shown that the privacy margin increases quadratically in the inverse of the angular separation between the true and the fake paths under Eve's perspective. Numerical simulations validate the theoretical findings on CRBs and showcase the approach's benefit in terms of bit error rates achievable by Bob and Eve.
\end{abstract}

\section{Introduction}

There is a growing interest in physical layer security in wireless communication as user interactions with critical services are becoming more ubiquitous.
Although novel approaches to protect users' location are to be developed, 
most are software or network-based and operate on the upper layers of the communication stack; see \emph{e.g.}~\cite{Farhang2015,Jiang2021} and references therein. They are, however, ineffective in securing the lower-level radio transmissions, which remain subject to leakage or eavesdropping of physical communication parameters, such as the channel state information and, thus, the end-user's locations.

To prevent location leakage, protocols
adding artificial pseudo-random noise to the transmitted signal or to spatially discriminate the physical space via beamforming~\cite{goel2008GuaranteeingSecrecy, Oh2012, Wang2015, Checa2020, tomasin2022BeamformingArtificial, Ayyalasomayajula} are proposed. In particular, Goel et al.~\cite{goel2008GuaranteeingSecrecy} achieve a secrecy capacity by introducing artificial noise into the null space of the legitimate receiver channel with or without the help of amplifying relays. The case of synchronized transmissions from cooperative agents to distort the measured received signal at the eavesdropper location in studied in~\cite{Oh2012}.
Artificial noise and beamforming are combined in~\cite{tomasin2022BeamformingArtificial} to minimize the signal power at unwanted stations. 
However, the aforementioned noise injection and beamforming approaches rely on implicit assumptions about the eavesdropper channel, which can be hard to verify in practice. Furthermore, those methods hide the transmitters' locations from all receiving parties and exclude the option to disclose those locations to legitimate receivers.

Of interest to this work, a novel beamforming scheme for location privacy via fake path injection was proposed  in~\cite{li2023ChannelState} in the context of multiple-input single-output orthogonal frequency division multiplexing (MISO-OFDM) communication. Unlike the previous, this method does not require channel information and allows the disclosure of the transmitter's location to legitimate parties. A key in the privacy analysis of this scheme is the stability of the \emph{line spectral estimation} problem (\emph{a.k.a} super-resolution), which consists of recovering the frequencies of complex exponentials from a finite number of  samples~\cite{kay1981spectrum}.

\subsection{Contributions and Organization of the Paper}
Inspired by~\cite{li2023ChannelState}, we consider the problem of protecting the location of a transmitter (Alice) from an eavesdropper (Eve) in her communication with a legitimate receiver (Bob). We focus on the unchartered case where the communication medium is a single-input multiple-output (SIMO) geometric multi-path channel. When Bob and Eve operate uniform linear antenna arrays (ULA), both parties can attempt to estimate the angle-of-arrivals (AoAs), leaking information on Alice's location.
To ease Bob's channel estimation and create a channel identifiability gap between Bob and Eve, it is assumed that Alice and Bob can exchange fake path parameters over a secure side channel at a \emph{low information rate.} In this setting, it is established that fake path injection can degrade Eve's estimation of the true AoAs and the channel coefficients, inducing privacy.

To assess the privacy benefit of the communication paradigm, a novel \emph{statistical privacy margin} is introduced in Definition~\ref{def:stat_secrecy} as the quotient between the smallest eigenvalue of Eve's Cramér-Rao lower bound (CRB) on the channel parameters based with the largest eigenvalue of Bob's one. This statistical metric is more stringent than that adopted in~\cite{li2023ChannelState}, as extremal eigenvalues of the CRB \emph{uniformly} control the error estimate of \emph{any linear combination} of the channel parameters. Exploiting recent mathematical advances in characterizing the condition number of \emph{generalized Vandermonde matrices}~\cite{AUBEL2019,Batenkov2020,Kunis2021colliding,Ferreira2023Higher,Ferreira2023Higher}, we show the proposed privacy metric is essentially driven by two key channel characteristics: 1) the minimal separation of the true paths; 2) the separation between fake and true paths.

The rest of this paper is organized as follows. The SIMO communication model with side information channel and the secrecy metric conducting our analysis are presented in Section~\ref{sec:problem_statement}, where their relevance is discussed. Section~\ref{sec:main_results} presents our main privacy results. Two cases are considered:
\begin{enumerate}
    \item Eve knows the channel coefficients and only infers Alice's AoAs;
    \item Eve infers both the channel coefficients and Alice's AoAs.
\end{enumerate}
In both cases, explicit bounds on the secrecy margin are provided. In particular, the importance of small angular separations between the true and the fake paths in Eve's perspective in inducing secrecy is highlighted. Section~\ref{sec:proof} presents the proof of the preceding results. Numerical simulations validating the theoretical secrecy bounds and showcasing the privacy benefits of fake-path injection in terms of Bob's and Eve's bit error rates (BERs) are provided in Section~\ref{sec:numerical_simulations}, and a conclusion is drawn in Section~\ref{sec:conclusion}.

\subsection{Mathematical Notation and Definitions}
Let $\bbT = \bbR/\bbZ$ be the unidimensional torus. For any $\tau \in \bbT$, we define $\bv_0(\tau) \in \bbC^N$ and $\bv_1(\tau) = \frac{\mathrm{d}\bv_0(\tau)}{\mathrm{d}\tau} \in  \bbC^N$ as,
\begin{subequations}
	\begin{align}
		\bv_0(\tau) ={}& \frac{1}{\sqrt{N}}\left[e^{-i 2 \pi (-n) \tau}, \dots, e^{i 2 \pi n \tau}\right]^\top \\
		\bv_1(\tau) ={}& \frac{1}{\sqrt{N}} \left[i 2 \pi (-n) e^{i 2 \pi (-n) \tau}, \dots, i 2 \pi n e^{i 2 \pi n \tau}\right]^\top,
	\end{align}
\end{subequations}
where $2n + 1 = N$.
For any set $\btau = \{ \tau_1, \dots, \tau_L \} \subset \bbT$ of $L$ elements, we define by $\bV_0(\btau), \bV_1(\btau) \in \bbC^{N \times L}$, the \emph{generalized Vandermonde matrices}
	\begin{align}\label{eq:V-def}
		\bV_p(\btau) ={}& \left[ \bv_p(\tau_1), \dots, \bv_p(\tau_L) \right], \quad p \in \{0,1\}.
	\end{align}
 The normalized Dirichlet kernel of order $N=2n+1$, written $D_N$, is given for any $t\in \bbR$ by $D_N(t)=\frac{1}{N}\sum_{k=-n}^{n} e^{-i 2\pi k t}$, which is an infinitely derivable, periodic function with period $1$. Of particular interest are the identities $D_N(0) = 1$, \mbox{$D_N^\pprime(0) = - \frac{\pi^2}{3}(N-1)(N+1)$}, $D_N^{(4)}(0) = \frac{\pi^4}{5} N^4(1+o(1))$.
 
\section{Problem Statement}\label{sec:problem_statement}
\subsection{SIMO Model with Fake Path Injection}\label{subsec:SIMO_model}
    We consider localization in SIMO communication, in which Eve overhears the transmission between Alice and Bob. We assume both Bob and Eve operate ULAs of $N_B$ and $N_E$ antennas, respectively. Classically, the ULAs are assumed of half-wavelength spacing $d = \frac{\lambda}{2}$, where $\lambda$ is the carrier wavelength. Additionally, we assume both the Alice--Bob and the Alice--Eve channels are linear with $L$ users (or $L$-paths). We write $\bc_B = \{c_{B,1},\dots,c_{B,L} \} \subset \bbC$ and
    $\bc_E = \{c_{E,1},\dots,c_{E,L} \} \subset \mathbb{C}$ Bob's and Eve's channel coefficients, respectively,  while $\btau_B = \{\tau_{B,1},\dots,\tau_{B,L} \} \subset \bbT$ and $\btau_E =\{\tau_{E,1},\dots,\tau_{E,L} \}$ encode Bob's and Eve's AoAs $\phi_{B,\ell}, \, \phi_{E,\ell} \in [-\frac{\pi}{2}, \frac{\pi}{2})$ through the relation $\tau = \frac{d \sin(\phi)}{\lambda} \in \bbT$. Furthermore, the pilot sequence transmitted by Alice is assumed to be known to Bob and Eve.

    To obscure her physical location to Eve when sending pilots, Alice emits \emph{fake paths}---carrying no relevant information--- which are modeled at the receivers as spurious paths parameterized by coefficients $\bm{\tilde{c}}$ and angular mapping $\bm{\tilde{\tau}}$. The received signals $\bm{y}_{B} \in \mathbb{C}^{N_B}$ and $\bm{y}_{E} \in \mathbb{C}^{N_{E}}$ at Bob's and Eve's side, write, respectively,
    \begin{subequations}\label{eq:model}
    \begin{align}
        \bm{y}_{B} = \mathbf{V}_0(\bm{\tau}_{B}) \bm{c}_{B} + \mathbf{V}_0(\bm{\tilde{\tau}}_{B}) \bm{\tilde{c}_{B}} + \bm{w}_{B},\\
        \bm{y}_{E} = \mathbf{V}_0(\bm{\tau}_{E}) \bm{c}_{E} + \mathbf{V}_0(\bm{\tilde{\tau}}_{E}) \bm{\tilde{c}_{E}} + \bm{w}_{E},
    \end{align} 
    \end{subequations}
    where $\bm{w}_{B} \sim \mathcal{CN}(\mathbf{0}, \eta^2_{B} \mathbf{I}_{N_B})$ and $\bm{w}_E \sim \mathcal{CN}(\mathbf{0}, \eta^2_{E} \mathbf{I}_{N_E})$ are noises.
    In the sequel, the subscripts ``B'' or ``E'' are dropped for notation convenience unless explicit disambiguation is needed.
    In practice, fake paths can emerge from Alice's beamforming design~\cite{li2023ChannelState} or a third-party jammer cooperating with Alice.
    
\begin{figure}[t]
    \vspace{2pt}
    \centering
    \includegraphics[width=0.92\linewidth]{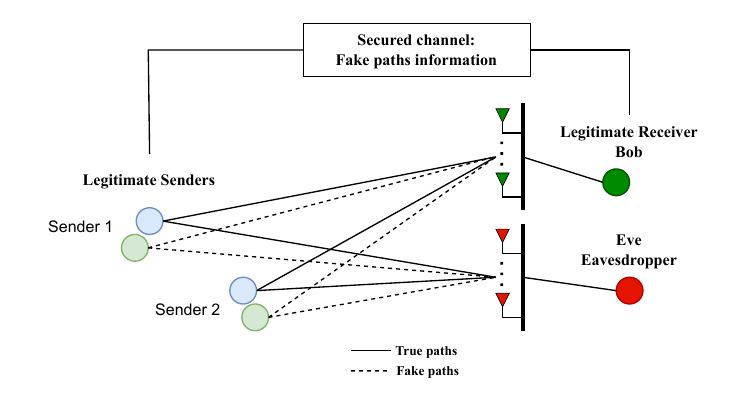}
        \caption{SIMO communication model with fake path injection.}
        \label{fig:sys_model}
        \vspace{-6pt}
    \end{figure}

    To resolve the ambiguities introduced by the fake signal design, we assume Alice and Bob can privately exchange securely at a \emph{low communication rate} through a side channel that is not overheard by Eve. The studied setup assumes the sharing of the artificial fake paths' parameters $\{\bctil, \btautil\}$. In practice, those parameters can be pre-agreed before the transmission or coded, such as in the context of a wiretap channel with side information~\cite{oggier2011SecrecyCapacity}. Hence, Bob can remove the faked component from its received signal before estimating the true signal parameters, while Eve is left to estimate both fake and true channel components, which is statistically harder. 
    The system model is depicted in Figure~\ref{fig:sys_model}.

    \subsection{Statistical Privacy Metric}

    Alice's physical location can be inferred by estimating the AoA's $\{\phi_\ell\}$, or equivalently the $\tau_\ell$'s from the observation $\by$ given by Equation~\eqref{eq:model}. As the columns of $\bV_0(\btau)$ in~\eqref{eq:V-def} are complex exponential vectors, estimating the model parameters amounts to solving a \emph{line spectral estimation} problem~\cite{kay1981spectrum,krim1996two}, which is a fundamental signal processing primitive. The privacy of Alice's location can be assessed by Eve's achievable estimation error in estimating the parameters $\{\bc, \btau\}$.
    
    In this work, we measure privacy in terms of the eigenvalue gap between  Bob's and Eve's Cramér-Rao Bound (CRB) matrices. Our privacy margin is formally defined as follows.
    \begin{definition}[Statistical privacy margin]
        \label{def:stat_secrecy}
        Assume Bob and Eve attempt to estimate an unknown parameter $\btheta$ from an observation $\bm{y}$ and write $\mathbf{CRB}$ the CRB matrix on $\btheta$ under the observation $\by$. Given $\gamma \geq 1$, is it said that \emph{statistical privacy} is achieved with margin $\gamma$ when
        \begin{equation}\label{eq:stat_secrecy}
            \frac{\lambda_{\min} (\mathbf{CRB}_{\mathrm{E}}(\btheta_E))}{\lambda_{\max} (\mathbf{CRB}_{\mathrm{B}}(\btheta_B))} \geq \gamma .
        \end{equation}
    \end{definition}
    Statistical privacy with margin $\gamma$ implies the quadratic error of any unbiased estimator in Eve's attempt to estimate \emph{any linear form of the parameters $\btheta_E$} is at least $\gamma$-times greater than Bob's best-achievable error on \emph{any other linear form of the parameters $\btheta_B$}. Hence,  Definition~\ref{def:stat_secrecy} of statistical privacy is more stringent than requesting control of the quadratic error on one single parameter, which is proposed in~\cite{li2023ChannelState}.
    Furthermore, when the noise is Gaussian, the CRB is equivalent to the mean-squared error of the maximum likelihood (ML) estimator~\cite{stoica1990music}. In that case, statistical privacy with margin $\gamma$ indicates Bob's ML estimation error under a signal-to-noise-ratio (SNR) $\textsf{SNR}_B$ would be \emph{smaller} than that of Eve's provided $\textsf{SNR}_E \leq \gamma \textsf{SNR}_B$. Therefore, larger values of the privacy margin $\gamma$ imply greater privacy on the parameters~$\btheta$.

    \section{Main Results}\label{sec:main_results}
    This section provides statistical privacy guarantees on the model presented in Section~\ref{sec:problem_statement} through the lens of Definition~\ref{def:stat_secrecy}. We start by introducing the \textit{wrap-around distance} $\Delta$, defined as the minimal distance over the torus between two distinct elements in $\btau$, and the \emph{inter-separation} $\delta$, defined as Hausdorff distance between the true and the paths:
    \begin{subequations}\label{eq:separations}
        \begin{align}
\Delta = \Delta(\btau) & \triangleq \min_{\ell \neq \ell^\prime} \inf_{j \in \bbZ} \left\vert \tau_\ell - \tau_{\ell^\prime} + j\right\vert\label{eq:min-sep} \\
        \delta = \delta(\btau, \btautil) & \triangleq \max_{\ell, \ellp} \inf_{j\in \bbZ} \left\vert \tau_\ell - \tilde{\tau}_{\ell^\prime} + j \right\vert.\label{eq:fake-sep}
\end{align}
    \end{subequations}
The above metrics play a critical role in the stability of line spectrum estimation and the estimator's consistency~\cite{Ankur,da2018tight}.

In the sequel, statistical privacy is studied for two distinct scenarios. First, we study the CRB on $\btau$ under the assumption that Bob and Eve know the channel coefficients $\{\bc, \bctil \}$, corresponding to a mono-static scenario where the channel true paths are invariant. Second, we consider the more generic setting under the hypothesis that Bob and Eve are agnostic of any channel parameters and study their CRB on $\{\btau, \bc\}$.

    \subsection{Privacy on the Angle of Arrivals}\label{subsec:aoa}

    We assume a simplified scenario where the channel coefficients $\{\bc, \bctil\}$ are known to both Bob and Eve. Hence, Bob's only remaining unknown is $\{\btau \}$, while Eve's unknowns are $\{ \btau, \btautil \}$. We highlight that knowledge of the channel coefficients is \emph{favorable} to Eve, and the current modeling is thus conservative for a privacy study. Theorem~\ref{theo:privacy_aoa} presents desirable bounds on the extremal eigenvalues of Bob and Eve's CRB on the parameter $\btau$ given measurement $\by$ of the form~\eqref{eq:model}.
    \begin{theorem}\label{theo:privacy_aoa}
    Assume $\Delta_B \geq \frac{\pi^2}{N}$, and $\delta_E < \frac{\Delta_E}{2}$. Then there exist two numerical constants $C > 0$ and $C'>0$ such that
    \begin{subequations}
        \begin{align}
        \lambda_{\max} (\mathbf{CRB}_{\mathrm{B}}(\btau_B))  \hspace{-2pt} &\leq  \hspace{-2pt}\frac{\pi^2}{3} \eta^2_B N_B^2{\left(1 - \pi^2\left(N_B\Delta_B \right)^{-1} \right)}^{-1} & \label{eq:privacy_aoa_bob}   \\
        \lambda_{\min} (\mathbf{CRB}_{\mathrm{E}}(\btau_E)) \hspace{-2pt}& \geq  \hspace{-2pt} \frac{\eta_E^2 N_E^4}{4\delta^{2}_E} \hspace{-3pt}\left( C   \hspace{-2pt}+ \hspace{-2pt}\frac{C'   \log \left( \frac{L}{2} \right)}{ N_E\Delta_E \left(1 \hspace{-3pt} - \hspace{-3pt}  \frac{2\delta}{\Delta_E} \right)} \hspace{-3pt} \right)^{\hspace{-12pt} -1} \hspace{-9pt} 
        \label{eq:privacy_aoa_eve}
    \end{align}
    \end{subequations}
    \end{theorem}
    Taking the quotient between the two quantities~\eqref{eq:privacy_aoa_eve} and~\eqref{eq:privacy_aoa_bob} immediately yields Corollary~\ref{cor:privacy_margin_aoa}.
    \begin{corollary}\label{cor:privacy_margin_aoa}
        Under the assumption of Theorem~\ref{theo:privacy_aoa} the SIMO communication model described in Section~\ref{sec:problem_statement} is statistically private in the sense of Definition~\ref{def:stat_secrecy} with secrecy margin
        \begin{equation}
            \gamma = \frac{3\eta^2_{E}}{4\pi^2\eta^2_{B}} \frac{N_E^4}{N_B^2 \delta_B^2}\frac{1 - \pi^2 {(N_B \Delta_B)}^{-1}}{ C   + \frac{C'   \log \left( \frac{L}{2} \right)}{ N_E\Delta_E \left(1 -   \frac{2\delta}{\Delta_E} \right)}  },
        \end{equation}
    where $C > 0$ and $C'>0$ are the constants of Theorem~\ref{theo:privacy_aoa}.
    \end{corollary}
    Corollary~\ref{cor:privacy_margin_aoa} indicates that provided Bob perceives well-separated paths, the secrecy margin increases with the ratio between Eve's and Bob's noise levels and decreases as the distance between the true and the fake paths increases. Additionally, increasing the number of antennas with $N = N_B = N_E$ while maintaining a constant separation factor $N\delta_E$ yields a polynomial increment in the privacy margin $\gamma = \mathcal{O}(N^4)$.
    
    \subsection{Channel Coefficients--AoAs Privacy}
    Herein is considered the generic case where both AoA and channel parameters are unknown.
    Since the vectors $\bc$ and $\btau$ are of different units, and since the statistical error of an estimator of $\btau$ scales inversely proportional to $N$ \cite{Costa23StabilitySR}, we apply the normalization $u_\ell =  \sqrt{-D_N^\pprime (0)} \tau_\ell$ and control the CRB on the set of parameter $\btheta = \{\bc, \bu \}$ instead. Bounds on the CRBs and the privacy margin are given in Theorem~\ref{theo:privacy_aoa_channel} and Corollary~\ref{cor:privacy_margin_aoa_channel}.

    \begin{theorem}\label{theo:privacy_aoa_channel}
    Assume $\Delta_B \geq \frac{\pi^2}{N}$, and $\delta_E < \frac{\Delta_E}{2}$. Then there exist two constants $C, C^\prime > 0$ such that
    \begin{subequations}
        \begin{align}
        \MoveEqLeft[1] \lambda_{\max} (\mathbf{CRB}_{\mathrm{B}}(\btheta_B)) \leq \eta^2_B {\left(1 - \pi^2\left(N_B\Delta_B \right)^{-1} \right)}^{-1} & \label{eq:privacy_all_bob}  \\
        \MoveEqLeft[1] \lambda_{\min} (\mathbf{CRB}_{\mathrm{E}}(\btheta_E)) \geq \frac{\eta^2_E N_E^2}{4\delta^{2}_E }\hspace{-3pt} \left( C   \hspace{-2pt} + \hspace{-2pt} \frac{C^\prime \log \left( \frac{L}{2} \right)}{ N_E\Delta_E \left(1 - \frac{2\delta_E}{\Delta_E} \right)}  \right)^{-1} \hspace{-5pt}. \label{eq:privacy_all_eve}
    \end{align}
    \end{subequations}
    \end{theorem}

     \begin{corollary}\label{cor:privacy_margin_aoa_channel}
        Under the assumption of Theorem~\ref{theo:privacy_aoa_channel} the SIMO communication model described in Section~\ref{sec:problem_statement} is statistically private in the sense of Definition~\ref{def:stat_secrecy} with secrecy margin
        \begin{equation}
            \gamma \geq \frac{\eta^2_E}{4\eta^2_B} \frac{N_E^4}{N_B^2 \delta_B^2}  \frac{1 - \pi^2 {(N_B \Delta_B)}^{-1}}{ C   + \frac{C'   \log \left( \frac{L}{2} \right)}{ N_E\Delta_E \left(1 -   \frac{2\delta}{\Delta_E} \right)} },
        \end{equation}
        where $C,C^\prime > 0$ are the numerical constants of Theorem~\ref{theo:privacy_aoa_channel}.
    \end{corollary}
    The trends on $\gamma$ proposed by Corollary~\ref{cor:privacy_margin_aoa_channel} is similar to that of Corollary~\ref{cor:privacy_margin_aoa}, when channel coefficients are known.
    
\section{Proofs of Theorem~\ref{theo:privacy_aoa} and Theorem~\ref{theo:privacy_aoa_channel}}
\label{sec:proof}
In this section, we demonstrate the main results. Preliminary bounds on the Dirichlet kernel are given in~\ref{subsec:dirichlet}, then Theorems~\ref{theo:privacy_aoa} and \ref{theo:privacy_aoa_channel} are demonstrated in Sections~\ref{subsec:proof_aoa} and~\ref{subsec:proof_aoa_channel}.

\subsection{Numerical Bounds on the Dirichlet Kernel}\label{subsec:dirichlet}
 The next lemma proposes bounds for the infinite norm of matrices involving the Dirichlet kernel in their generic terms.

\begin{lemma}\label{lem:dirichlet_numerical_bounds}
    Let $\btau \subset \bbT$ and $\btautil \subset \bbT$ two sets of cardinality $L$, with maximal inter-separation $\delta$ as in~\eqref{eq:fake-sep}. Let $\bG_p$ the matrix with generic term, for $p \in \{0,1,2\}$,%\vspace{-10pt}
\begin{multline}\label{eq:Gp_def}
        \bG_p(i,j) = \\
        D_N^{(p)}(\tau_i - \tau_j) - D_N^{(p)}(\tautil_i - \tau_j) + D_N^{(p)}(\tautil_i - \tautil_j) - D_N^{(p)}(\tau_i - \tautil_j).
\end{multline}
If $\delta < \frac{\Delta(\btau)}{2}$, then there exist constants $C_p \geq 0$ such that
\begin{equation}
    \norm{\bG_p}_\infty \leq 4 \delta^2 \left(\sup_{|\varepsilon| \leq 2\delta}\left\vert D_N^{(p+2)}(\varepsilon) \right\vert  + \frac{C_p N^{p+2} \log \left( \frac{L}{2} \right)}{ N\Delta(\btau) \left(1 - \frac{2\delta}{\Delta(\btau)} \right)}  \right).
\end{equation}
\end{lemma}
\vspace{-8pt}
%\addtolength{\topmargin}{9mm}
\begin{IEEEproof}
First of all, Taylor's expansion and the mean value theorem with the assumption $|\tau_\ell- \tautil_\ell| \leq \delta$ yields
\begin{align}\label{eq:bound_Gp_delta}
    \left\vert \bG_p(i,j) \right\vert 
    & \leq 4 \delta^2 \sup_{|\varepsilon| \leq 2\delta} \left\vert D_N^{(p+2)} \left( \tau_i - \tau_j + \varepsilon \right) \right\vert.
\end{align}
\vspace{-6pt}
Next, we claim the existence of constants $C_p \geq 0$ such that
\begin{equation}\label{eq:bound_dirichlet}
    \left\vert D_N^{(p+2)}(t) \right\vert \leq C_p N^{p+1} \left\vert t \right\vert^{-1}, \quad \forall t \in [-\frac{1}{2}, \frac{1}{2}),\; p \in \{0,1,2\}.
\end{equation}
Similar bounds are studied in~\cite{Kunis2021colliding} for the derivatives up to the second order. Herein, we claim~\eqref{eq:bound_dirichlet} holds with the constants $C_0 = 5, C_1=16, C_2=50$. We fix the value of $i$. From the separation condition~\eqref{eq:min-sep}, one can reorder the indexes without loss of generality, so that
\(0\leq |j-i| \Delta(\btau) \leq |\tau_i - \tau_j| \leq 1
\) for all $j$.
Therefore, Equations~\eqref{eq:bound_Gp_delta} and \eqref{eq:bound_dirichlet}, and the decreasing of $|t|^{-1}$  over $[0,\frac{1}{2})$ induce
\begingroup
\allowdisplaybreaks
\begin{align}\label{eq:bound_sum_Fij}
    \MoveEqLeft[0] \sum^L_{j=1} |\bG_p(i,j)| = |\bG_p(i,i)| + \sum_{\substack{1 \leq j \leq L  \\ j \neq i }} |\bG_p(i,j)| & \nonumber \\
    & \leq 4 \delta^2 \hspace{-3pt} \left( \hspace{-3pt} \sup_{|\varepsilon| \leq 2\delta}\left\vert D_N^{(p+2)}(\varepsilon) \right\vert   + \hspace{-8pt} \sum_{\substack{1 \leq j \leq L  \\ j \neq i }} \sup_{|\varepsilon| \leq 2\delta} \hspace{-2pt}\left\vert D_N^{(p+2)} \left( \tau_i - \tau_j + \varepsilon \right) \right\vert \hspace{-3pt}\right)
    \nonumber \\
    & \leq 4 \delta^2 \left( \sup_{|\varepsilon| \leq 2\delta}\left\vert D_N^{(p+2)}(\varepsilon) \right\vert + \sum_{\substack{1 \leq j \leq L  \\ j \neq i }} \sup_{|\varepsilon| \leq 2\delta} \frac{C_p N^{p+1}}{ \left\vert \tau_i - \tau_j + \varepsilon \right\vert} \right) \nonumber \\
    & \leq 4 \delta^2 \left( \sup_{|\varepsilon| \leq 2\delta}\left\vert D_N^{(p+2)}(\varepsilon) \right\vert + 2 C_p N^{p+1} \sum_{k=1}^{\left\lceil \frac{L-1}{2} \right\rceil}  \frac{1}{ k \Delta - 2\delta  } \right).
\end{align}
\endgroup
As the bound~\eqref{eq:bound_sum_Fij} holds independently of $i$, one may conclude on the desired result with the identity $\sum_{k=1}^{\left\lceil \frac{L-1}{2} \right\rceil}  \frac{1}{ k \Delta - 2\delta  } \leq \frac{1}{2 \Delta(\btau) \left(1 - \frac{2\delta}{\Delta(\btau)} \right)}  \log \left( \frac{L}{2} \right) \leq$ for $\delta < \frac{\Delta(\btau)}{2}$.
\end{IEEEproof}

\subsection{Proof of Theorem~\ref{theo:privacy_aoa}}\label{subsec:proof_aoa}

We start by proving the bound~\eqref{eq:privacy_aoa_bob} for Bob's estimation of $\btau$. As the fake paths parameters $\{\bctil, \btautil \}$ are shared by Alice to Bob, and as the channel coefficients $\bc$ are assumed to be known, Bob only needs to estimate $\btau$. The Fisher information matrix (FIM) writes
\begin{align}
    \bJ_{B} (\btau) = \eta^{-2} \diag(\bc)^\herm \bV_1(\btau)^\herm \bV_1(\btau) \diag(\bc),
\end{align}
which immediately implies through the relation $\mathbf{CRB}_{\mathrm{B}}(\btau) = {\bJ_{B}(\btau)}^{-1}$, and with~\cite[Thm. 4]{Ferreira2023Higher}, \cite{Costa23Condtion}
\begin{align}
    \lambda_{\max} (\mathbf{CRB}_{\mathrm{B}}(\btau)) &= \lambda_{\min} {(\bJ_{B}(\btau))}^{-1} \nonumber \\
    & \leq \eta^2 |c_{\min}|^{-2} \lambda_{\min} \left(\bV_1(\btau)^\herm \bV_1(\btau) \right)^{-1} \nonumber \\
    & \leq \eta^2 |c_{\min}|^{-2} \frac{\pi^2}{3}N^2 \left(1 - \pi^2 (N\Delta)^{-1} \right)^{-1}.
\end{align}

Next, we demonstrate Eve's estimation bound~\eqref{eq:privacy_aoa_eve}. 
As Eve is assumed to know the channel coefficients, her unknowns in the observation model~\eqref{eq:model} reduces to $\btheta = \{\btau, \tilde{\btau} \}$, and her FIM on $\btheta$ under the observation $\by$ writes~\cite[Chapter~5]{VanTreesDetection}
\begin{align}\label{eq:FIM_aoa}
    \MoveEqLeft[0.5] \bJ_{E} (\btheta) = & \nonumber \\
    &\eta^{-2}  \diag(\bc, \tilde{\bc})^\herm
        \begin{bmatrix}
           \bV_1(\bm{\tau})^\herm \bV_1(\bm{\tau}) & \bV_1(\bm{\tau})^\herm \bV_1(\bm{\tilde{\tau}}) \\
            \bV_1(\bm{\tilde{\tau}})^\herm \bV_1(\bm{\tau}) & \bV_1(\bm{\tilde{\tau}})^\herm \bV_1(\bm{\tilde{\tau}})
        \end{bmatrix}\diag(\bc, \tilde{\bc}).
\end{align}
The CRB matrix is given as the inverse of~\eqref{eq:FIM_aoa}. Schur's inversion formula yields~\cite{SCHARF1993geometry}
\begin{align}
    \mathbf{CRB}_{\mathrm{E}} (\btheta) 
    \hspace{-2pt} &= \hspace{-2pt} \eta^{2}\hspace{-1pt} \diag(\bc, \tilde{\bc})^{-1}
    \hspace{-1pt}\begin{bmatrix}
            \bM^{-1} &  \hspace{-3pt}* \\
            * &  \hspace{-3pt}\tilde{\bM}^{-1}
        \end{bmatrix}\hspace{-1pt}
        \diag(\bc, \tilde{\bc})^{-\herm}
\end{align}
with $\bM = \bV_1(\btau)^\herm \bP_1^\perp(\btautil) \bV_1(\btau)$, $\tilde{\bM} = \bV_1(\btautil)^\herm \bP_1^\perp(\btau) \bV_1(\btautil)$, where $\bP_1^\perp(\btau)$ and $\tilde{\bP}_1^\perp(\btautil)$ are the projection matrices onto the orthogonal complement of the column space of $\bV_1(\btau)$ and $\bV_1(\btautil)$, respectively. Hence, Eve's CRB matrix on the AoA relevant parameters $\btau$ satisfies
\begin{align}\label{eq:CRB_expression_AoA}
    \lambda_{\min}\left(\mathbf{CRB}_{\mathrm{E}}(\btau)\right) &= \lambda_{\min} \left( \eta^2 \diag(\bc)^{-\herm} \bM^{-1}   \diag(\bc)^{-1} \right) \nonumber \\
    &\geq \eta^2 |c_{\max}|^{-2} {\lambda_{\max}\left( \bM \right)}^{-1}
\end{align}
It remains to provide an upper bound on  $\lambda_{\max} \left( \bM \right)$ to conclude. We let $\bV_1(\btau) = \bV_1(\btautil) + \bE$. The expression of $\bM$ reduces to
\begin{equation}
    \bM = \bE^\herm \bP_1^\perp(\btau) \bE.
\end{equation}
Next, by a direct calculation of the generic term $\bE^\herm \bE$ reveals the identify $\bG_2 = \bE^\herm \bE$, where $\bG_2$ is as in~\eqref{eq:Gp_def}. Hence, by the contractivity of the orthogonal projection $\bP_1^\perp(\btau)$ one as
\begin{align}\label{eq:bound_M_AoA}
    \lambda_{\max}(\bM) & \leq \lambda_{\max}(\bE^\herm \bE) = \lambda_{\max}(\bG_2) \leq \norm{\bG_2}_{\infty}.
\end{align}
Combining~\eqref{eq:CRB_expression_AoA} with~\eqref{eq:bound_M_AoA} and applying Lemma~\ref{lem:dirichlet_numerical_bounds} concludes on~Inequation \eqref{eq:privacy_aoa_eve}.\hfill \IEEEQEDhere

\subsection{Proof of Theorem~\ref{theo:privacy_aoa_channel}}\label{subsec:proof_aoa_channel}
\label{sec:secure_cc_AoA}
First of all, we define by $\bW(\btau)$ the concatenation $\bW(\btau) = \left[\bV_0(\tau), \frac{1}{\sqrt{-D_N^\pprime(0)}}\bV_1(\btau) \right]$. We structure the proof analogously to that of Theorem~\ref{theo:privacy_aoa}.

We start by considering Bob's case, for whom the unknowns are $\btheta = \{\bc, \bu \}$. The FIM writes
\begin{equation}
\label{eqn:all_J_Bob}
    \bfJ_{B} (\btheta) = \eta^{-2} \diag(\bm{1}, \bc)^\herm \bW(\btau)^\herm \bW(\btau) \diag(\bm{1}, \bc),
\end{equation}
and one establishes~\eqref{eq:privacy_all_bob} with~\cite[Theorem 4]{Ferreira2023Higher}, ~\cite{Costa23Condtion} with
\begin{align}
    \lambda_{\max} (\mathbf{CRB}_{\mathrm{B}}(\btheta)) &= \lambda_{\min} {(\bJ_{B}(\btau))}^{-1} \nonumber \\
    & \leq \eta^2 |c_{\min}|^{-2} \lambda_{\min} \left(\bW(\btau)^\herm \bW(\btau) \right)^{-1} \nonumber \\
    & \leq \eta^2 |c_{\min}|^{-2}  \left(1 - \pi^2 (N\Delta)^{-1} \right)^{-1}.
\end{align}

As for Eve, her unknowns are $\bar{\btheta} = \{\bc, \bu , \bctil, \butil \}$, and her CRB matrix writes~\cite[Chapter~5]{VanTreesDetection}
\begin{align}
    \mathbf{CRB}_{E} (\bar{\btheta}) \hspace{-1pt}
    &= \hspace{-1pt} \eta^{2} {\diag(\bc, \tilde{\bc})}^{-1} \hspace{-1pt}
    \begin{bmatrix}
            \bN^{-1} &  \hspace{-3pt}* \\
            * &  \hspace{-3pt}\tilde{\bN}^{-1}
        \end{bmatrix}\hspace{-1pt}
        {\diag(\bc, \tilde{\bc})}^{-\herm}
\end{align}
with $\bN = \bW(\btau)^\herm \tilde{\bQ}^\perp \bW(\btau)$ and $\tilde{\bN} = \bW(\btautil)^\herm \tilde{\bQ}^\perp \bW(\btautil)$, and where $\bQ^\perp$ and $\tilde{\bQ}^\perp$ are the projection matrices onto the orthogonal complement of the column space of $\bW(\btau)$ and $\bW(\btautil)$, respectively. Hence,Eve's CRB on the partial set of relevant parameter $\btheta = \{\bc, \bu\}$ can be lower bounded by
\begin{align}\label{eq:CRB_expression_1}
    \lambda_{\min}\left(\mathbf{CRB}_{\mathrm{E}}(\btheta)\right) &= \lambda_{\min} \left( \eta^2 \diag(\bc)^{-1} \bN^{-1}   \diag(\bc)^{-
    \herm} \right) \nonumber \\
    &\geq \eta^2 |c_{\max}|^{-2} {\lambda_{\max}\left( \bN \right)}^{-1}.
\end{align}
It remains to upper bounding $\lambda_{\max}\left( \bN \right)$ to conclude. We let $\bW(\btau) = \tilde{\bW}(\btautil) + \bF$. The expression of $\bN$ reduces to
\begin{equation}
    \bN = \bF^\herm \tilde{\bQ}^\perp \bF.
\end{equation}
A direct calculation of the generic term of $\bF^\herm \bF$ yields the block decomposition
\begin{equation}
    \bF^\herm \bF = \begin{bmatrix}
        \bG_0 & \frac{1}{\sqrt{-D_N^\pprime(0)}}\bG_1 \\
        \frac{1}{\sqrt{-D_N^\pprime(0)}}\bG_1^\herm & -\frac{1}{D_N^\pprime(0)}\bG_2
    \end{bmatrix}
\end{equation}
where $\bG_p$ are the matrices defined in~\eqref{eq:Gp_def},  $p\in \{0,1,2\}$. Furthermore, $\bQ^\perp$ is an orthogonal projection, hence is contractive.
\begin{align}\label{eq:bound_M}
    \lambda_{\max}(\bN)  &\leq \lambda_{\max}(\bF^\herm \bF) \leq \norm{\bF^\herm \bF}_\infty \nonumber \\
    &\leq \max \left\{ \norm{\bG_0}_\infty + \frac{1}{\sqrt{-D_N^\pprime(0)}}\norm{\bG_1}_\infty , \right. \nonumber \\
    & \qquad \left. \frac{1}{\sqrt{-D_N^\pprime(0)}}\norm{\bG_1}_\infty - \frac{1}{D_N^\pprime(0)} \norm{\bG_2}_\infty  \right\}.
\end{align}
Combining~\eqref{eq:CRB_expression_1} with~\eqref{eq:bound_M} and applying Lemma~\ref{lem:dirichlet_numerical_bounds} concludes on~Inequation \eqref{eq:privacy_aoa_eve}.\hfill \IEEEQEDhere

\section{Numerical Experiments}\label{sec:numerical_simulations}

\subsection{Validation of the theorems}\label{subsec:validation}

We provide numerical insight on the results presented in Section~\ref{sec:main_results}. We assume Bob and Eve have $N_B=N_E=31$ antennas, $L=5$ true paths with equispaced AoAs for both Bob and Eve in the parameter space, that is $\Delta_B = \Delta_E = 1/L$. The SNR is defined as $\mathsf{SNR} = \norm{\bV_0(\btau)}^2_2 / \norm{\bw}^2_2$, and we set $\mathsf{SNR}_B = \mathsf{SNR}_E = 0\mathrm{dB}$. We consider the scenario described in Section~\ref{subsec:aoa}, where the channel coefficients are assumed to be known to both Bob and Eve.

\begin{figure}[t]
    %\vspace{-8pt}
    \centering
    \includegraphics[width=0.9\linewidth]{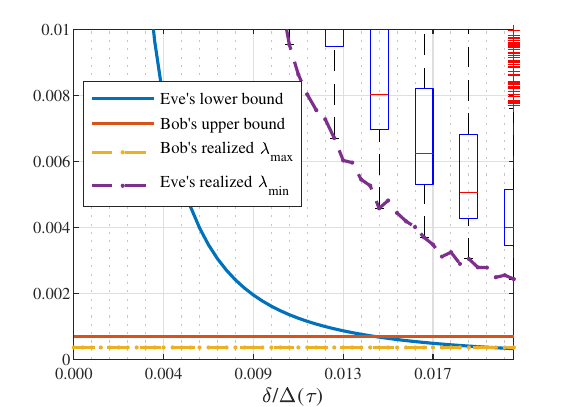}
        \vspace{-6pt}
    \caption{Theoretical and realized extremal values of Bob's and Eve's CRB, case of known channel coefficients.}
    \label{fig:AoA_bound}
\end{figure}

Figure~\ref{fig:AoA_bound} pictures the theoretical bounds on Bob's and Eve's CRBs established in Theorem~\ref{theo:privacy_aoa} as a function of the ratio $\delta_E / \Delta_E$. A comparison is made with the numerical realization of the CRB for the random realization of the fake paths while ensuring $|\btautil_\ell - \btau_\ell| \leq \delta$ of all paths $\ell \in \{1, \dots, L\}$. The histogram of the empirical realization of $\lambda_{\min} (\mathbf{CRB}_{\mathrm{E}}(\btheta_E))$ as well as its extremal values are displayed. Also, there is a gap between the theoretical and realized bound on Eve's CRB---possibly due to the coarse majoration in Lemma~\ref{lem:dirichlet_numerical_bounds}---, the trend is captured by our theoretical predictions. 

\begin{figure}
\vspace{-12pt}
    \centering     \includegraphics[width=0.9\linewidth]{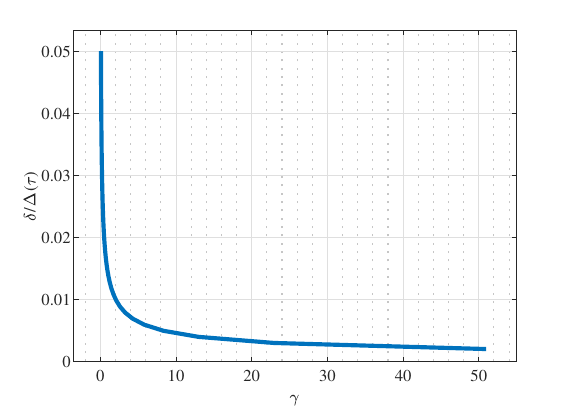}
    \vspace{-6pt}
        \caption{Fake path separation needed to achieve a target secrecy margin $\gamma$, case of known channel coefficients.}
        \label{fig:AoA_secrecy_margin}
    \vspace{-12pt}
\end{figure}

Figure~\ref{fig:AoA_secrecy_margin} shows the ratio $\frac{\delta_E}{\Delta_E}$ that is requested to achieve a given secrecy margin $\gamma$, which corroborates with Corollary~\ref{cor:privacy_margin_aoa}.


\subsection{Physical layer security for multi-user SIMO communication}

In this experiment, we consider a BPSK-modulated SIMO communication with $L=3$ users, while Bob's and Eve's base stations (BS) are equipped with $N_B = N_E =15$ antennas. Each user has a direct line-of-sight to both BSs, and a cooperative jammer transmits the fake paths. The communication is split into two steps: 1) the users transmit 15 pilot symbols, and Bob and Eve employ the MUSIC algorithm to estimate the AoAs and the channel coefficients; 2) the users transmit unknown messages, and Bob and Eve perform maximum likelihood decoding to determine the messages. True paths are equispaced so that $\Delta = 1/L$, while fake paths are fixed to
$\tautil_i = \tau_i + \delta$.

\begin{figure}
     \centering
     \begin{subfigure}[b]{0.9\linewidth}
         \centering
         \includegraphics[trim={15pt 4pt 30pt 5pt},clip,width=\textwidth]{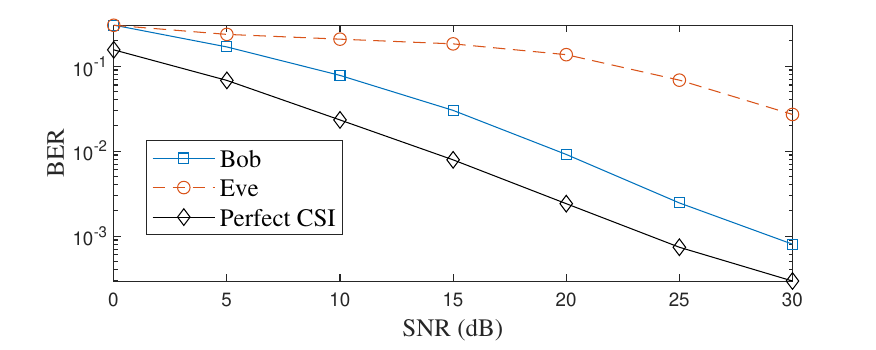}
         \caption{$\delta / \Delta = 10^{-1}$}
         \label{fig:ber-ratio-10}
     \end{subfigure}
     
     \begin{subfigure}[b]{0.9\linewidth}
         \centering
         \includegraphics[trim={15pt 4pt 30pt 5pt},clip,width=\textwidth]{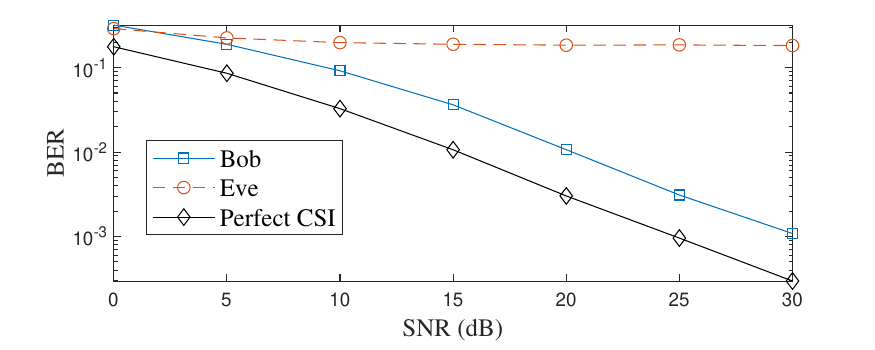}
         \caption{$\delta / \Delta = 10^{-2}$}
         \label{fig:ber-ratio-200}
     \end{subfigure}
        \caption{BER of Bob and Eve after estimating the CSI}
        \label{fig:ber}
        \vspace{-12pt}
\end{figure}

Figure~\ref{fig:ber} draws the BERs achieved by Bob and Eve for two different values of the ratio $\delta/\Delta$. Decoding under perfect CSI is shown in a black line as a baseline for our experiment. We note from the graphs that the SNR gap between Bob's BER and decoding under perfect CSI remains approximately constant at around 6 dB w.r.t the change in SNR. Meanwhile, Eve's BER is greater than Bob's, and the gap between Eve's BER and the one achieved with perfect CSI increases as the SNR increases. Eve's performance further degrades as $\delta$ decreases, corroborating with  Theorem~\ref{theo:privacy_aoa_channel} and with the numerical evaluation of the secrecy margin conducted in Section~\ref{subsec:validation}.

\section{Conclusion}\label{sec:conclusion}
In this work, we proposed a novel scheme enhancing the location privacy of transmitters in a SIMO communication paradigm by injecting fake paths, whose parameters are secretly shared between Alice and Bob over a secure side channel. Privacy is assessed in our framework under a novel statistical privacy metric based on the extremal eigenvalues of Bob's and Eve's CRB matrices on the true path parameters. Theoretical guarantees back the privacy enhancements, which mainly depend on the angular distance between the true and the fake paths under Eve's perspective. We leave for future work a generalization of the privacy framework proposed in Definition~\ref{def:stat_secrecy} to more general contexts, such as the study of location privacy in MIMO communication.

\IEEEtriggeratref{4}
\renewcommand*{\bibfont}{\footnotesize}
\printbibliography

\end{document}